\documentstyle[twocolumn,aps]{revtex}
\input{epsf}
\newcommand{\eff}{\mbox{\scriptsize\em eff}}

\begin{document}
\draft

\title{Nonfactorization in Cabibbo-Favored $B$ Decays}
\author{F. M. Al-Shamali and A. N. Kamal}
\address{Theoretical Physics Institute and Department of Physics,
University of Alberta, Edmonton, Alberta T6G 2J1, Canada. \\
\vspace{1.5mm} {\em Report no.\ : Alberta Thy 12-98 \\ hep-ph/9806270}}

\maketitle

\begin{abstract}
We assume universal values for the color-singlet ($\varepsilon_1$) and color octet ($\varepsilon_8$) nonfactorization parameters in $B$ Decays.  Two sets of color-favored processes and one set of color-suppressed processes were used to give quantitative estimates of these parameters. It has been found (by calculating the branching ratios for a large number of Cabibbo-favored $B$ Decays) that the values $\varepsilon_1(\mu_0) = - 0.06 \pm 0.03$ and $\varepsilon_8(\mu_0) = 0.12 \pm 0.02$ improve significantly the predictions of the factorization model.
\end{abstract}



\section{Introduction}
The idea of nonfactorization in $D$ and $B$ decays was introduced in the last few years by several authors \cite{ref:Cheng-94,ref:Kamal-94,ref:Soares-95}. There are two equivalent ways of introducing nonfactorized contributions in a calculation: Either, use the number of colors, $N_c$ equal to 3 and explicitly add a nonfactorized contribution to each Lorentz scalar in the decay amplitude as in \cite{ref:Kamal-94,ref:Soares-95}, or introduce an effective number of colors, $N^{eff}_c$. The later approach has been adopted in several papers \cite{ref:Ali-97,ref:Ali-98,ref:Cheng-98} dealing with $B$ decays into light mesons.

Though the nonfactorized amplitude remains incalculable, a few statements about it can be made. First, what is estimated to be the nonfactorized contribution depends on the model of form factors used to calculate the factorized contribution. Second, within a chosen model for the form factors, the nonfactorized contributions are process dependent. Third, if nonfactorization is characterized through an effective number of colors, $N^{eff}_c$, is it the same (as assumed in \cite{ref:Ali-97} and \cite{ref:Ali-98}) for the tree and the penguin generated processes or different (as in \cite{ref:Cheng-98})? Ref.~\cite{ref:Cheng-98} goes so far as to suggest that $N^{eff}_c$, is different for the $(V - A) \, (V - A)$ and $(V - A) \, (V + A)$ type of penguin terms. In the language of \cite{ref:Kamal-94}, where $N_c = 3$ is assumed, different $N^{eff}_c$, for the tree and penguin diagram driven processes imply that the nonfactorized contributions for the tree and penguin driven diagram processes are different. This is not unlikely as the calculation of the penguin diagram driven amplitudes in the factorization assumption involves additional assumptions, an effective value of $q^2$ for example. In a complete calculation \cite{ref:Kamal-97} there are no such free momentum choices left.

Unlike the papers listed in \cite{ref:Ali-97,ref:Ali-98,ref:Cheng-98}, in this paper we study $B$ decays into a heavy and a light meson. As such they are all Cabibbo-favored $b \rightarrow c$ transitions. Following, \cite{ref:Shamali-98} we use $N_c = 3$ and introduce nonfactorized contributions explicitly. In order to evaluate the factorized contributions we use the Bauer-Stech-Wirbel (BSW) model \cite{ref:Wirble-85} to calculate the form factors at $q^2 = 0$ but extrapolate according to the BSW II prescription [see the discussion following Eq.~(\ref{eq:I-VV}) for details].

To achieve simplicity of description, we assume that the nonfactorized effects associated with the three Lorentz-scalar structures in $B \rightarrow V V$ decays are the same. Using experimental data, we then calculate the average overall nonfactorization factors for color-favored $b \rightarrow c \bar{u} d$ and $b \rightarrow c \bar{c} s$ processes. We repeat this for color-suppressed $b \rightarrow c \bar{c} s$ processes. We relate these nonfactorization factors to the scale dependent parameters $\varepsilon_1(\mu)$ and $\varepsilon_8(\mu)$ of \cite{ref:Neubert-97} and determine them at $\mu_0 = 4.6$ GeV. The details are explained in the text of this paper. Finally, having isolated the parameters $\varepsilon_1(\mu_0)$ and $\varepsilon_8(\mu_0)$, we make predictions for the decay rates of Cabibbo-favored modes measured and as yet unmeasured. The measured rates are shown to agree well with the predictions  with few exceptions.

The paper is arranged as follows Section II deals with the formalism and introduces notations. Sections III and IV deal with the calculation of nonfactorized amplitudes, the evaluation of $\varepsilon_1(\mu_0)$ and $\varepsilon_8(\mu_0)$, and the predictions of branching ratios. A discussion of the results and conclusions appear in Sec.\ V\@.


\section{Formalism}
\subsection{Effective Hamiltonian}
In the absence of strong interactions, the effective Hamiltonian for the process
$b \rightarrow c \bar{u} d$ is given by
\begin{equation}
{\cal H}_{\mbox{\scriptsize eff}} =  \frac{G_F}{\sqrt{2}} V_{cb} V_{ud}^*
\, ( \bar{c} b )_L \, ( \bar{d} u )_L ,
\label{eq:Heff-b-cud}
\end{equation}
where
\begin{eqnarray}
(\bar{c} b)_L & = & \bar{c}_i \gamma^\mu (1 - \gamma_5) b_i , \nonumber\\
(\bar{d} u)_L & = & \bar{d}_i \gamma^\mu (1 - \gamma_5) u_i ,
\label{eq:V-Acurrent}
\end{eqnarray}
and $i$ is the color index. When QCD effects are included, the effective Hamiltonian is generalized to~\cite{ref:Buras-92,ref:Buchalla-96}
\begin{equation}
{\cal H}_{\mbox{\scriptsize eff}} =  \frac{G_F}{\sqrt{2}} V_{cb} V_{ud}^*
\left[ C_1 \; (\bar{c} b)_L \, (\bar{d} u)_L + C_2 \, (\bar{c} u)_L \; (\bar{d} b)_L \right] ,
\label{eq:Heff-QCD1}
\end{equation}
where $(\bar{c} b)_L \, (\bar{d} u)_L$ and  $ (\bar{c} u)_L \, (\bar{d} b)_L$ are current $\times$ current  local operators.

The Wilson coefficients, $C_1$ and $C_2$, include the short-distance QCD corrections. Their values depend on the renormalization scale $\mu$ through the following renormalization group equation (RGE)
\begin{equation}
\mu \, \frac{d C_\pm}{d \mu} = \frac{\alpha_s}{4 \pi} \, \gamma_\pm \, C_\pm ,
\label{eq:Cpm-DE}
\end{equation}
where
\begin{equation}
C_\pm = \frac{1}{\sqrt{2}} (C_2 \pm C_1) 
\end{equation}
and
\begin{equation}
\alpha_s (\mu) = \frac{4 \pi}{\beta_0 \ln (\mu^2/\Lambda^2)} .
\end{equation}
In the leading order the coefficient $\beta_0$ is given by
\begin{equation}
\beta_0 = \frac{1}{3} ( 11 \, N_c - 2 \, N_f  ) ,
\end{equation}
where $N_c$ is the number of colors and $N_f$ is the number of active flavors. The eigenvalues of the anomalous dimension matrix are
\begin{equation}
\gamma_- = - 8, \hspace{1.5cm}  \gamma_+ = 4\; .
\end{equation}

When Eq.\ (\ref{eq:Cpm-DE}) is solved it gives
\begin{equation}
C_\pm (\mu) = \left( \frac{\alpha_s (m_W)}{\alpha_s (\mu)} \right)^{\gamma_\pm/2 \beta_0} C_\pm (m_W) .
\end{equation}
At the particular scale ($\mu_0 = 4.6 \; \mbox{GeV} \sim m_b$) we have
\begin{eqnarray}
C_1 & = & \;\;\;1.127 \pm 0.005 \nonumber\\
C_2 & = & - 0.286 \pm 0.008 \; . \label{eq:C1-C2}
\end{eqnarray}
For the QCD scale we used the value
$\Lambda^5_{\overline{\mbox{\scriptsize MS}}} = 219 \pm 24$ MeV \cite{ref:PDG-98} which corresponds to $\alpha (m_Z) = 0.119 \pm 0.002$.
The uncertainties in $C_1$ and $C_2$ are due to the uncertainties in
$\Lambda^5_{\overline{\mbox{\scriptsize MS}}} $.

Similarly, the other Cabibbo-favored process ($b \rightarrow c \bar{c} s$) occurs through the Hamiltonian
\begin{equation}
{\cal H}_{\mbox{\scriptsize eff}} =  \frac{G_F}{\sqrt{2}} V_{cb} V_{cs}^*
\left[ C_1 \; (\bar{c} b)_L \, (\bar{s} c)_L + C_2 \, (\bar{c} c)_L \; (\bar{s} b)_L \right] \; ,
\label{eq:Heff-QCD2}
\end{equation} 
with the same values for the Wilson coefficients as in (\ref{eq:C1-C2}).


\subsection{Factorization and Nonfactorization}
Consider the color-favored decay $\overline{B}^0 \rightarrow D^+ \pi^-$.
The decay amplitude for this process is given by
\begin{eqnarray}
&& {\cal A}(\overline{B}^0 \rightarrow D^+ \pi^-) = \langle D^+ \pi^- |
{\cal H}_{\mbox{\scriptsize eff}} | \overline{B}^0 \rangle \nonumber\\
&& \qquad = \frac{G_F}{\sqrt{2}} V_{cb} V_{ud}^*
\left[ \rule{0.0mm}{6.0mm} a_1 \, \langle D^+ \pi^- | (\bar{c} b) \, (\bar{d} u) | \overline{B}^0 \rangle \right. \nonumber\\
&& \qquad + \left. C_2 \, \langle D^+ \pi^- | \, \frac{1}{2} \sum_a (\bar{c} \lambda^a b)
\, (\bar{d} \lambda^a u) \, | \overline{B}^0 \rangle \right] ,
\label{eq:amplitude-PP1}
\end{eqnarray}
where Fierz transformation and color-algebra have been used to rearrange the quark flavors.
In (\ref{eq:amplitude-PP1})
\begin{equation}
a_1(\mu) = \left(C_1(\mu) + \frac{C_2(\mu)}{N_c} \right) ,
\end{equation}
where $\lambda^a$ are the Gell-Mann matrices.

The second term on the right hand side of (\ref{eq:amplitude-PP1}) is nonfactorizable, while the first term receives both factorizable and nonfactorizable contributions. So we write \cite{ref:Cheng-94,ref:Kamal-94,ref:Soares-95,ref:Neubert-97}
\begin{eqnarray}
&& {\cal A} (\overline{B}^0 \rightarrow D^+ \pi^-) = \frac{G_F}{\sqrt{2}} V_{cb} V_{ud}^* \, a_1
\nonumber\\
&& \quad \times \left[ \rule{0.0mm}{6.0mm} \langle D^+ |  (\bar{c} b) | \overline{B}^0 \rangle \langle \pi^- |  (\bar{d} u) | 0 \rangle\,
+ \langle D^+ \pi^- | (\bar{c} b) \, (\bar{d} u) | \overline{B}^0 \rangle^{nf} \right. \nonumber\\
&& \quad + \left. \frac{C_2}{a_1} \, \langle D^+ \pi^- | \, \frac{1}{2} \sum_a (\bar{c} \lambda^a b) \, (\bar{d} \lambda^a u) \, | \overline{B}^0 \rangle^{nf} \right] .
\label{eq:amplitude-PP2}
\end{eqnarray}
Following the conventions in \cite{ref:Neubert-97}, we define the following color-singlet and color-octet nonfactorization parameters:
\begin{eqnarray}
\varepsilon^{(BD, \pi)}_1 & = & \frac{\langle D^+ \pi^- | (\bar{c} b) \, (\bar{d} u) | \overline{B}^0 \rangle^{nf}}{\langle D^+ |  (\bar{c} b) | \overline{B}^0 \rangle \langle \pi^- |  (\bar{d} u) | 0 \rangle}
, \label{eq:epsilon1}\\
\varepsilon^{(BD, \pi)}_8 & = & \frac{\langle D^+ \pi^- |  \, \frac{1}{2} \sum_a (\bar{c} \lambda^a b) \, (\bar{d} \lambda^a u) \, | \overline{B}^0 \rangle^{nf}}{\langle D^+ |  (\bar{c} b) | \overline{B}^0 \rangle \langle \pi^- |  (\bar{d} u) | 0 \rangle} .
\label{eq:epsilon8}
\end{eqnarray}
The decay amplitude then takes the form
\begin{eqnarray}
{\cal A} (\overline{B}^0 \rightarrow D^+ \pi^-) = \frac{G_F}{\sqrt{2}} V_{cb} V_{ud}^* \,
a_1 \xi^{(BD, \pi)}_1 \nonumber\\
\times \; \langle D^+ |  (\bar{c} b) | \overline{B}^0 \rangle \langle \pi^- | (\bar{d} u) | 0 \rangle ,
\label{eq:amplitude-PP-I}
\end{eqnarray}
where
\begin{equation}
\xi^{(BD, \pi)}_1(\mu) = \left( 1 + \varepsilon^{(BD, \pi)}_1(\mu) +
\frac{C_2}{a_1} \, \varepsilon^{(BD, \pi)}_8(\mu) \right) .
\end{equation}

If we consider the color-suppressed decay $\overline{B}^0 \rightarrow D^0 \pi^0$,
the effective Hamiltonian (\ref{eq:Heff-QCD1}) is rewritten, using Fierz transformations, as
\begin{equation}
{\cal H}_{\mbox{\scriptsize eff}} = \frac{G_F}{\sqrt{2}} V_{cb} V_{ud}^*
\left[ a_2 \, (\bar{c} u) \, (\bar{d} b) +
C_1 \, \frac{1}{2} \sum_a (\bar{c} \lambda^a u) \, (\bar{d} \lambda^a b) \right] ,
\label{eq:Heff-QCD-CS-1}
\end{equation}
where
\begin{equation}
a_2(\mu) = \left(C_2(\mu) + \frac{C_1(\mu)}{N_c} \right) .
\end{equation}
The decay amplitude, then, takes the form
\begin{eqnarray}
{\cal A} (\overline{B}^0 \rightarrow D^0 \pi^0) = \frac{G_F}{\sqrt{2}} V_{cb} V_{ud}^* \, a_2 \, \xi^{(B\pi, D)}_2\, \nonumber\\
\times \; \langle \pi^0 | (\bar{d} b) | \overline{B}^0 \rangle \langle D^0 | (\bar{c} u) | 0 \rangle ,
\label{eq:amplitude-PP-II}
\end{eqnarray}
where
\begin{equation}
\xi^{(B\pi, D)}_2(\mu) = \left( 1 + \varepsilon^{(B\pi, D)}_1(\mu) +
\frac{C_1}{a_2} \, \varepsilon^{(B\pi, D)}_8(\mu) \right) .
\end{equation}

In this work, we will assume that $\varepsilon_1$ and $\varepsilon_8$ are universal constants for all Cabibbo-favored $B$ decays. For this reason, their superscripts will be dropped from now on.

The decay amplitudes for the other processes of class I (color-favored) and processes of class II (color-suppressed) considered in this work, are derived from (\ref{eq:amplitude-PP-I}) and (\ref{eq:amplitude-PP-II}), respectively, by making appropriate replacements. However, for processes of class III, which receive contributions from both $a_1$ and $a_2$, we use a suitable combination of the above mentioned equations to derive their amplitude. For example, the decay amplitude for the process $B^- \rightarrow D^0 \pi^-$ is given by
\begin{eqnarray}
&& {\cal A} (B^- \rightarrow D^0 \pi^-) = \frac{G_F}{\sqrt{2}} V_{cb} V_{ud}^* \nonumber\\
&& \qquad \times \left[ \, a_1 \, \xi_1 \; \langle D^0 |  (\bar{c} b) | B^- \rangle \langle \pi^- | (\bar{d} u) | 0 \rangle \right. \nonumber\\
&& \qquad + \; \left. a_2 \, \xi_2 \; \langle \pi^- | (\bar{d} b) | B^- \rangle \langle D^0 | (\bar{c} u) | 0 \rangle \, \right] \; .
\label{eq:amplitude-PP-III}
\end{eqnarray}

Equations (\ref{eq:amplitude-PP-I}) and (\ref{eq:amplitude-PP-II}) suggest the following definitions for the effective $a_1$ and $a_2$
\begin{eqnarray}
a^{\eff}_1 & = & a_1 \, \xi_1 = a_1 \left[ 1 + \varepsilon_1 \right] +
C_2 \, \varepsilon_8 , \\
a^{\eff}_2 & = & a_2 \, \xi_2 = a_2 \left[ 1 + \varepsilon_1 \right] +
C_1 \, \varepsilon_8 .
\end{eqnarray}
The coefficients $a^{\eff}_1$ and $a^{\eff}_2$ are independent of the renormalization scale since the $\mu$ dependence of the Wilson coefficients is compensated by the $\mu$ dependence of nonfactorization parameters. So, the RGE in (\ref{eq:Cpm-DE}) leads to \cite{ref:Neubert-97}
\begin{eqnarray}
\varepsilon_1(\mu) & = & \frac{1}{2} \left[ \left( 1 + \frac{1}{N_c} \right) \left[ 1 + \varepsilon_1(\mu_0) \right] + \varepsilon_8(\mu_0) \right] \nonumber\\
&& \times \left[ \frac{\alpha_s(\mu)}{\alpha_s(\mu_0)} \right]^{\gamma_+/2 \beta_0} \nonumber\\
& + & \frac{1}{2} \left[ \left( 1 - \frac{1}{N_c} \right) \left[ 1 + \varepsilon_1(\mu_0) \right] - \varepsilon_8(\mu_0) \right] \nonumber\\
&& \times \left[ \frac{\alpha_s(\mu)}{\alpha_s(\mu_0)} \right]^{\gamma_-/2 \beta_0} - 1
\label{eq:E1} \\
\varepsilon_8(\mu) & = & \frac{1}{2} \left[ \left( 1 - \frac{1}{N_c^2} \right) \left[ 1 + \varepsilon_1(\mu_0) \right] + \left( 1 - \frac{1}{N_c} \right) \varepsilon_8(\mu_0) \right]
\nonumber\\
&& \times \left[ \frac{\alpha_s(\mu)}{\alpha_s(\mu_0)} \right]^{\gamma_+/2 \beta_0} \nonumber\\
& - & \frac{1}{2} \left[ \left( 1 - \frac{1}{N_c^2} \right) \left[ 1 + \varepsilon_1(\mu_0) \right] -
\left(1 + \frac{1}{N_c} \right) \varepsilon_8(\mu_0) \right] \nonumber\\
&& \times \left[ \frac{\alpha_s(\mu)}{\alpha_s(\mu_0)} \right]^{\gamma_-/2 \beta_0} ,
\label{eq:E8}
\end{eqnarray}
where $\mu_0$ is arbitrary.

From $1/N_c$ expansion \cite{ref:Neubert-97,ref:Witten-79,ref:Buras-86} it is found that
\begin{eqnarray}
\varepsilon_1(\mu) & = & {\cal O}(1/N_c^2) \; , \nonumber\\
\varepsilon_8(\mu) & = & {\cal O}(1/N_c) \; . \label{eq:x1-x8-Nc}
\end{eqnarray}
A simple way to see this is to realize that whereas only one gluon exchange is needed to cause color-octet current to couple to color-singlet hadrons, two gluons are needed in the case of color-singlet currents.


\subsection{Current Matrix Elements}
Let $\left| I \right>$ and $\left| P \right>$ be pseudoscalar mesons and $\left| V \right>$ be a vector meson.  The hadronic current matrix elements can be decomposed in terms  of form factors and decay constants using Lorentz invariance. We define \cite{ref:Wirble-85}
\begin{eqnarray}
\langle P |  J_\mu | 0 \rangle & = & f_P \, (p_P)_\mu, \label{eq:MED.a} \\
\langle V |  J_\mu | 0 \rangle & = & m_V f_V \epsilon^*_\mu, \label{eq:MED.b} \\
\langle P |  J_\mu | I \rangle & = & \left( p_I + p_P -
\frac{m^2_I - m^2_P}{q^2} q \right)_\mu F_1(q^2) \nonumber\\
&& + \; \frac{m^2_I - m^2_P}{q^2} q_\mu \, F_0(q^2) , \label{eq:MED.c} \\
\langle V |  J_\mu | I \rangle & = &  \frac{- 2 i} { m_I + m_V} \varepsilon_{\mu\nu\rho\sigma} \epsilon^{*\nu} p^\rho_I \, p^\sigma_V \, V(q^2) \nonumber \\
& & + \; (m_I + m_V ) \epsilon^*_\mu A_1(q^2) \nonumber\\
&& - \; \frac{\epsilon^*.q}{ m_I + m_{V} }  (p_I + p_{V})_\mu \, A_2(q^2) \nonumber \\
& & - \; 2 m_{V} \frac{\epsilon^*.q}{ q^2} q_\mu \, \left[A_3(q^2) - A_0(q^2) \right] \label{eq:MED.d}
\end{eqnarray}
with
\begin{eqnarray}
q_\mu & = & \left( p_I - p_{P (V)} \right)_\mu  , \\
A_3(q^2) & = & \frac{m_I + m_V}{2 m_V} A_1(q^2) - \frac{m_I - m_V}{2 m_V} A_2(q^2) ,
\end{eqnarray}
and
\begin{eqnarray}
F_1(0) & = & F_0(0), \\
A_1(0) & = & A_3(0) .
\end{eqnarray}

The factorized current $\times$ current matrix elements, needed to calculate the decay amplitudes,  are evaluated using the above decomposition to be
\begin{eqnarray}
\langle P_1 |  J^\mu | I \rangle \, \langle P_2 |  J'_\mu | 0 \rangle & = & ( m^2_I - m^2_{P_1} ) f_{P_2} \, F_0(q^2) , \label{eq:I-PP}\\
\langle P |  J^\mu | I \rangle \, \langle V |  J'_\mu | 0 \rangle & = &
2 m_V f_V (\epsilon^*.p_I  ) \, F_1(q^2) , \label{eq:I-PV}\\
\langle V |  J^\mu | I \rangle \, \langle P |  J'_\mu | 0 \rangle & = &
2 m_V f_P (\epsilon^*.p_I ) \,A_0(q^2) , \label{eq:I-VP}
\end{eqnarray}
and
\begin{eqnarray}
\langle V_1 |  J^\mu | I \rangle \, \langle V_2 |  J'_\mu | 0 \rangle & = &
- \frac{m_{V_2}}{m_I + m_{V_1}} f_{V_2} \nonumber\\
& \times & \left[ 2 i \, \varepsilon_{\mu\nu\rho\sigma} \epsilon_2^{*\mu} \epsilon_1^{*\nu} p^\rho_{V_2} \, p^\sigma_{V_1} \, V(q^2) \right. \nonumber\\
& & - \;  (m_I + m_{V_1})^2 (\epsilon^*_1. \epsilon^*_2) \, A_1(q^2) \nonumber\\
&& \left. + \; 2 (\epsilon^*_1. p_{V_2}) (\epsilon^*_2. p_{V_1}) \, A_2(q^2) \right] .
\label{eq:I-VV}
\end{eqnarray}
where $\epsilon_1$ and $\epsilon_2$ are the polarization vectors of $V_1$ and $V_2$ respectively. 

For the form factors, we use BSW model \cite{ref:Wirble-85} to calculate their values at zero momentum transfer. Then, we extrapolate to the desired momentum using a monopole form for the form factors $F_0$ and $A_1$ and a dipole form for the form factors $F_1$, $A_0$, $A_2$ and $V$ \cite{ref:Neubert-92}. This is sometimes referred to as BSW II model. The two states $| \eta \rangle$ and $| \eta' \rangle$ are treated in the same way as in \cite{ref:Shamali-98} where the mixing angle and wavefunction normalizations are properly taken care of.

Regarding the decay constants we adopt the following values in this work
\cite{ref:Neubert-97,ref:PDG-98,ref:Richman-97,ref:CLEO-98}:
\begin{eqnarray}
f_{\pi} & = & 130.7 \pm 0.37 \;\; \mbox{MeV} \nonumber\\
f_{\rho} & = & 207 \pm 1  \;\; \mbox{MeV} \nonumber\\
f_{a_1} & = & 228 \pm 10  \;\; \mbox{MeV} \nonumber \\ 
f_{J/\psi} & = & 405 \pm 14  \;\; \mbox{MeV} \nonumber \\
f_{\psi(2S)} & = & 282 \pm 14  \;\; \mbox{MeV} \\
f_D & = & 200 \pm 10 \% \;\; \mbox{MeV} \nonumber \\
f_{D^*} & = & 230 \pm 10 \% \;\; \mbox{MeV} \nonumber\\
f_{D_s} & = & 250 \pm 27 \;\; \mbox{MeV} \nonumber \\
f_{D^*_s} & = & 275 \pm 10 \% \;\; \mbox{MeV} \nonumber
\end{eqnarray}

Now, from (\ref{eq:amplitude-PP-I}), (\ref{eq:amplitude-PP-II}), (\ref{eq:amplitude-PP-III}) and
(\ref{eq:I-PP} - \ref{eq:I-VV}) we can calculate the two-body decay amplitudes ${\cal A}(B \rightarrow M_1 \; M_2)$ where $M_1$ and $M_2$ are the two mesons (pseudoscalar or vector) in a given spin state. As a result, the branching ratios are given by
\begin{equation}
{\cal B}(B \rightarrow M_1 M_2) =
\frac{| {\bf k}|}{8 \pi m^2_B} \, \left| {\cal A}(B \rightarrow M_1 \; M_2) \right|^2 \, \tau_B,
\label{eq:decay-rate}
\end{equation}
where
\begin{equation}
| {\bf k}| = \frac{\left[  ( m^2_B - m^2_1 - m^2_2)^2 - 4 m^2_1 m^2_2 \right]^{1/2}}{2 m_B} 
\end{equation}
is the momentum of the decay products in the $B$ rest frame.

For the case when both $M_1$ and $M_2$ are vector mesons, the decay amplitude can be written in terms of the three helicity amplitudes (${\cal A}_0$, ${\cal A}_+$ and
${\cal A}_-$). The longitudinal ($P_0$) and transverse ($P_+$ and $P_-$) polarizations are then defined as
\begin{eqnarray}
P_0 & = & \frac{ |{\cal A}_0|^2}{|{\cal A}_0|^2 + |{\cal A}_+|^2 + |{\cal A}_-|^2}
\label{eq:P0} \\
P_\pm & = & \frac{ |{\cal A}_\pm|^2}{|{\cal A}_0|^2 + |{\cal A}_+|^2 + |{\cal A}_-|^2} \; .
\label{eq:Ppm}
\end{eqnarray}


\section{Evaluation of Nonfactorization Contribution}
Equations (\ref{eq:E1}) and (\ref{eq:E8}) give the explicit dependence of $\varepsilon_1$ and $\varepsilon_8$ on the renormalization scale if the values of these parameters are known at a particular point (e.g.\ $\mu_0$). However, the $\mu$ dependence is cancelled by the $\mu$ dependence of the Wilson coefficients such that the decay amplitude is $\mu$ independent. Our goal in this section is to estimate the values of $\varepsilon_1(\mu_0)$ and $\varepsilon_8(\mu_0)$ using available experimental data on color-favored and color-suppressed channels.

\begin{figure}
\epsfxsize=3.4in  \centerline{\epsffile{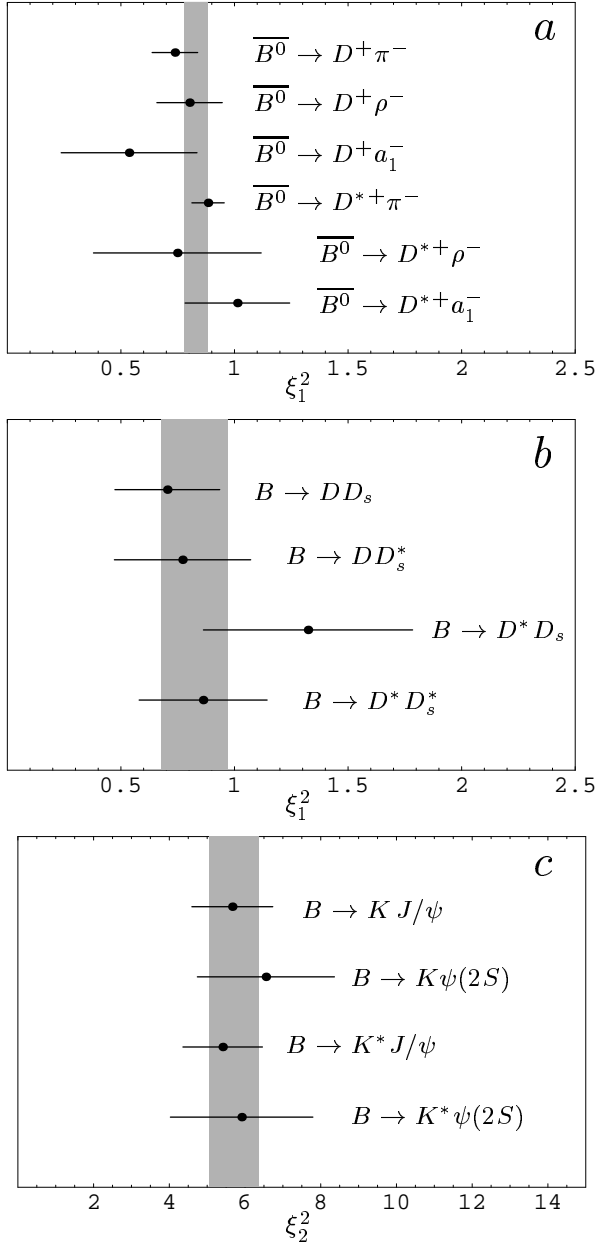}}

\caption{(a)The values of the nonfactorization parameter $\xi_1^2$ calculated for six color-favored processes of the type $b \rightarrow c \bar{u} d$. (b) The values of the nonfactorization parameter $\xi_1^2$ calculated for four color-favored processes of the type $b \rightarrow c \bar{c} s$. (c)The values of the nonfactorization parameter $\xi_2^2$ calculated for four color-suppressed processes of the type $b \rightarrow c \bar{c} s$. The shaded areas represent the statistical averages.}
\label{fig:xi2}
\end{figure}

Let us first consider the color-favored processes of the type $b \rightarrow c \bar{u} d$. By calculating the branching ratios of these decays using the factorization assumption ($\varepsilon_1 = \varepsilon_8 = 0$) and then by comparing it with experimental measurements we calculate $\xi^2_1$ in each decay channel using the simple relation
\begin{equation}
\xi^2 = \frac{{\cal B}(\mbox{experiment})}{{\cal B}(\mbox{factorization})} . \label{eq:xi2}
\end{equation}
The results are displayed in Fig.~\ref{fig:xi2}~(a).

We can see from these results that $\xi_1^2$ is almost channel independent except when the $a_1$ meson is present in the final state. By averaging the amount of nonfactorization in these channels we get the value, $\xi^2_1(\mu_0) = 0.83 \pm 0.05$. This corresponds to a $17 \%$ deviation from the factorization assumption.

The second set of processes considered, are the color-favored decays of the type $b \rightarrow c \bar{c} s$. The experimental value used for each decay mode were taken to be a weighted average over the charged and neutral decay channels. The resulting amount of nonfactorization in these decays are shown in Fig.~\ref{fig:xi2}~(b). The weighted average of the amount of nonfactorization in these processes, $\xi^2_1(\mu_0) = 0.82 \pm 0.15$, is in good agreement with that found in the decay processes of kind $b \rightarrow c \bar{u} d$.

Finally, we considered the color-suppressed processes of the type $b \rightarrow c \bar{c} s$. These processes give predictions regarding the nonfactorization parameter $\xi_2^2(\mu_0)$ using (\ref{eq:xi2}). In Fig.~\ref{fig:xi2}~(c) we show the results of these calculations. On the average, a value of $\xi^2_2(\mu_0) = 5.7 \pm 0.7$ is calculated.

\begin{figure}
\epsfxsize=3.4in  \centerline{\epsffile{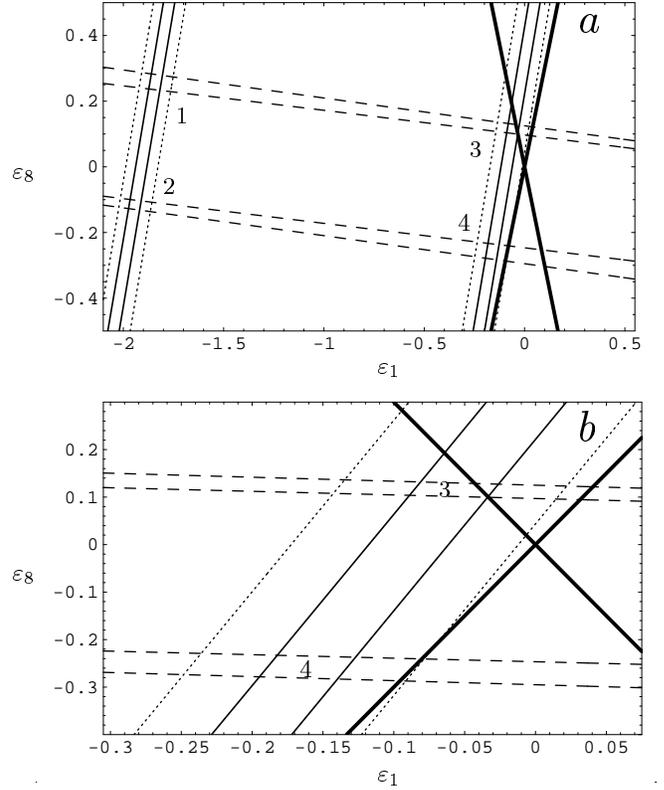}}

\caption{(a) The regions in $\varepsilon_1$-$\varepsilon_8$ space that correspond to the amount of nonfactorizations estimated for $\xi^2_1$ and $\xi^2_2$. The two parallel regions bounded by thin solid lines correspond to the value $\xi^2_1 = 0.83\pm 0.05$ calculated from the color-favored processes of the type $b \rightarrow c \bar{u} d$. The regions bounded by the dotted lines correspond to the value $\xi^2_1 = 0.82\pm 0.15$ calculated from the color-favored processes of the type $b \rightarrow c \bar{c} s$. The regions bounded by the dashed lines correspond to the value $\xi^2_2 = 5.7\pm 0.7$ calculated from the color-suppressed processes of the type $b \rightarrow c \bar{c} s$. The two intersecting solid lines correspond to the equation $\varepsilon_1 / \varepsilon_8 = \pm1/3$. (b) A magnification of the interesting region in $\varepsilon_1$-$\varepsilon_8$ space.}
\label{fig:eps18}
\end{figure}

Figure~\ref{fig:eps18}, shows the regions in $\varepsilon_1$-$\varepsilon_8$ space that correspond to the average values of $\xi^2_1$ and $\xi^2_2$ predicted above. In this figure, the two parallel regions bounded by thin solid lines correspond to the value, $\xi^2_1 = 0.83\pm 0.05$, calculated from the color-favored processes of the type $b \rightarrow c \bar{u} d$. The regions bounded by the dotted lines correspond to the value, $\xi^2_1 = 0.82\pm 0.15$, calculated from the color-favored processes of the type $b \rightarrow c \bar{c} s$. The value, $\xi^2_2 = 5.7\pm 0.7$, calculated from the color-suppressed processes of the type $b \rightarrow c \bar{c} s$ corresponds to the two horizontal bands bounded by dashed lines.

In Fig.~\ref{fig:eps18}, we also show two intersecting solid lines. These have been drawn in light of (\ref{eq:x1-x8-Nc}) and correspond to the equation
\begin{equation}
\frac{\varepsilon_1}{\varepsilon_8} = \pm \frac{1}{N_c} \,. \label{eq:x1-x8-Nc2}
\end{equation}
We realize that this relation need not be exact, nevertheless we use it to restrict our solutions.

As seen in Fig.~\ref{fig:eps18}, these regions intersect in four areas labeled 1, 2, 3 and 4. Areas 1 and 2 are excluded due to the severe violation of (\ref{eq:x1-x8-Nc2}). Area 4 is also excluded because it produces negative $a^{\eff}_2$ which is not supported by the mainstream data on $B$ decays. So, we are left with area 3. From the size and location of this area in the $\varepsilon_1$-$\varepsilon_8$ space, we get the following predictions for the color-singlet and color-octet nonfactorization parameters at $\mu_0$:
\begin{eqnarray}
\varepsilon_1(\mu_0) & = & - 0.06 \pm 0.03 , \nonumber\\
\varepsilon_8(\mu_0) & = & \;\;\> 0.12 \pm 0.02 \;. \label{eq:x1-x8-mu0}
\end{eqnarray}
These values correspond to
\begin{eqnarray}
\xi_1(\mu_0) = 0.91 \pm 0.03 & \;\; \longrightarrow \;\; & a^{\eff}_1 = 0.94 \pm 0.03 \, , \nonumber\\
\xi_2(\mu_0) = 2.39 \pm 0.23 & \;\; \longrightarrow \;\; & a^{\eff}_2 = 0.22 \pm 0.02 \;,
\label{eq:xi1-xi2-aeff}
\end{eqnarray}
and in the language of \cite{ref:Soares-95}, they correspond to
\begin{eqnarray}
\tilde{a}_1(\mu_0) & = & a_1^{\eff} - a_1(\mu_0) = - 0.09 \pm 0.03 \nonumber\\
\tilde{a}_2(\mu_0) & = & a_2^{\eff} - a_2(\mu_0) = \;\;\, 0.13 \pm 0.02 \;.
\end{eqnarray}
The estimated value of $\varepsilon_8$ is in agreement with that estimated in \cite{ref:Neubert-97} using a more limited set of processes.


\section{Predictions of Branching Ratios and Polarization}
Assuming universality of the nonfactorization parameters, $\varepsilon_1$ and $\varepsilon_8$, we calculated the branching ratios for a  large number of Cabibbo-favored decay channels which are shown  in Tables \ref{tab:B0-CF-bcud} - \ref{tab:Bs-CS-bccs}. The errors that appear in the theoretical calculations are due to the uncertainties in the Wilson coefficients, the decay constants, the $B$-meson lifetime and (where applicable) the mass of $a_1$ meson.

We start by studying six color-favored processes of the type $b \rightarrow c \bar{u} d$ shown in Table \ref{tab:B0-CF-bcud}. From this table, it can be seen that the inclusion of nonfactorization improves the predictions of the factorization assumption and produces a much better fit to experimental measurements. In Table \ref{tab:B0-CS-bcud}, we show the predicted branching ratios for ten color-suppressed processes of the type $b \rightarrow c \bar{u} d$. As the results show, the inclusion of nonfactorization in these decays enhances their decay rates, pushing them closer to the present upper bounds.

Tables \ref{tab:B-CF-bccs}  and \ref{tab:B-CS-bccs} show the predictions of the branching ratios for the color-favored and color-suppressed decays of the type $b \rightarrow c \bar{c} s$. By comparing these results with experimental data we find a significant improvement in the theoretical predictions, specially for the color-suppressed modes. The experimental value shown for each decay mode is a weighted average over the charged and neutral decay channels which differ in the flavor of their spectator quark. The average experimental values in \cite{ref:PDG-98} for the three processes $B \rightarrow K \psi(2S)$, $B \rightarrow K^* J/\psi$ and $B \rightarrow K^* \psi(2S)$ are updated to include the latest measurements \cite{ref:CDF-98} by CDF Collaboration. 

The four decay channels of $B^-$ meson shown in Table~\ref{tab:Bm-III-bcud} receive contributions from two diagrams. As a result the decay amplitudes contain interference between $a_1^{\eff}$ and $a_2^{\eff}$. Using the values of the nonfactorization parameters in (\ref{eq:x1-x8-mu0}) results in a constructive interference between the two diagrams and small modifications to the predictions of the factorization assumption. The predictions are in agreement with experimental results. However, if we used for the nonfactorization parameters the values $\varepsilon_1(\mu_0) = - 0.17$ and $\varepsilon_8(\mu_0) = - 0.26$, which are taken from area 4 in Fig.~\ref{fig:eps18}, the interference between the two diagrams becomes destructive. This results in lower values for the branching ratios and poor agreement with experiment supporting our choice of the solution in area 3 as discussed in the previous section.

Finally, we calculated the effect of nonfactorization  contribution on the branching ratios of three sets of $B_s$ decays. The results of these calculations are shown in Tables~\ref{tab:Bs-CF-bcud} - \ref{tab:Bs-CS-bccs}. Even though, experimental data on $B_s$-decays are very limited, the measured branching ratio of  $\overline{B}_s \rightarrow \phi J/\psi$ \cite{ref:PDG-98} shows encouraging agreement with our predictions.

In doing polarization calculations of color-favored and color-suppressed decays , the nonfactorization parameters cancel out. This is due to the assumed universality of $\varepsilon_1$ and $\varepsilon_8$. As a result, the polarization predictions are the same as in the factorization model though the branching ratios are scaled by an overall factor. The predictions of longitudinal ($P_0$) and transverse ($P_-$) polarizations of processes of the form $B \rightarrow VV$ are shown in Table~\ref{tab:B-Polarization}. Experimental measurements are available for a limited number of processes. For the decay $\overline{B}^0 \rightarrow D^{*+} \rho^-$ the available experimental value of $P_0$ is in agreement with the predicted value. In the case of $B \rightarrow K^* J/\psi$ decay the theoretical prediction of $P_0$ is about 30\% lower than the latest (and lowest) experimental measurement, and the predicted $P_-$ is about 35\% higher than the available experimental value. The process $\overline{B}_s \rightarrow \phi J/\psi$ also shows higher experimental value for $P_0$ than the predicted one. However, due to the limited data, it is difficult to make a strong statement about the validity of the factorization approach in predicting the polarizations of $B \rightarrow VV$ decays.

It is interesting to note that if we extrapolate all the form factors as monoples (the so called BSW II model) the predictions for the polarizations in $\overline{B}^0 \rightarrow D^{*+} \rho^-$, $B \rightarrow K^* J/\psi$ and $\overline{B}_s \rightarrow \phi J/\psi$ are in much better agreement with data, even though BSW II model fares much better in predicting the branching ratios. For the record, the longitudinal polarizations predicted in BSW I model for the above three decays are 0.87, 0.57 and 0.55 respectively


\section{Discussion and Conclusion}
Strictly speaking, we know that the factorization assumption can not be correct. This is because it produces a scale dependent transition amplitude. However, quoting authors of ref.\ \cite{ref:Neubert-97},
" what one may hope for is that it provides a useful approximation if the Wilson coefficients (or equivalently the QCD coefficients $a_1$ and $a_2$) are evaluated at a suitable scale $\mu_f$, the factorization point . "

In $B$ decays, the Wilson coefficients are usually evaluated at ($\mu \sim m_b$). If this is the factorization scale, then we should expect the nonfactorization parameters [$\varepsilon_1(m_b)$ and $\varepsilon_8(m_b)$] to vanish. However, since the predictions of the factorization assumption are generally not in agreement with data, specially for color-suppressed decays, this could indicate that the nonfactorizable parts of the decay amplitude are not negligible at the $B$-mass scale.

In this work, we tried to answer the following question: Is it possible via the introduction of  a minimum number of new parameters to improve the predictions of the factorization assumption and explain the bulk of available experimental data on Cabibbo-favored $B$ d$B$ decays?

In answering this question, we assumed universality of the color-singlet ($\varepsilon_1$) and the color-octet ($\varepsilon_8$) nonfactorization parameters. Two sets of color-favored processes and one set of color-suppressed processes were used to give quantitative estimates of these parameters. It has been found (by calculating the branching ratios for a large number of Cabibbo-favored $B$ decays) that the values $\varepsilon_1(\mu_0) = - 0.06 \pm 0.03$ and $\varepsilon_8(\mu_0) = 0.12 \pm 0.02$ improve significantly the predictions of the factorization assumption even though the Wilson coefficients were evaluated at leading order and only the tree diagrams were considered. These results support the argument that nonfactorization plays an important role in $B$ decays into two hadrons. 

\vspace{2.5mm}
\begin{center} \small\bf ACKNOWLEDGEMENTS \end{center}
This research was partially supported by a grant to A.N.K. from the Natural Sciences and  Engineering Research Council of Canada.



\begin{table}
\centering
\caption{The predicted branching ratios for the color-favored processes of the type $b \rightarrow c \bar{u} d$ calculated,  in column 2, by taking $\varepsilon_1(\mu_0) = \varepsilon_8(\mu_0) = 0$ and, in column 3, by taking $\varepsilon_1(\mu_0) = -0.06 \pm 0.03, \;\;\varepsilon_8(\mu_0) = 0.12 \pm 0.02 $. The last column represents the available experimental measurements}

\begin{tabular}{lccc} 
& \multicolumn{3}{c}{Branching Ratio $\times 10^{-3}$} \\ \cline{2-4}
Process & Fac. & Nonfac. & Exp. \cite{ref:PDG-98} \\\hline
$\overline{B}^0 \rightarrow D^+ \pi^-$ & $4.1 \pm 0.1$ & $3.4 \pm 0.3$ & $3.0 \pm 0.4$ \\
$\overline{B}^0 \rightarrow D^+ \rho^-$ & $9.9 \pm 0.3$ & $8.2 \pm 0.6$ & $7.9 \pm 1.4$ \\
$\overline{B}^0 \rightarrow D^+ a_1^-$ & $11.2 \pm 1.0$ & $9.3 \pm 1.1$ & $6.0 \pm 3.3$ \\
$\overline{B}^0 \rightarrow D^{*+} \pi^-$ & $3.1 \pm 0.1$ & $2.6 \pm 0.2$ &  $2.76 \pm 0.21$\\
$\overline{B}^0 \rightarrow D^{*+} \rho^-$ & $9.0 \pm 0.3$ & $7.4 \pm 0.6$ & $6.7 \pm 3.3$\\
$\overline{B}^0 \rightarrow D^{*+} a_1^-$ & $12.8 \pm 1.2$ & $10.6 \pm 1.2$  &  $13.0 \pm 2.7$\\
\end{tabular}

\label{tab:B0-CF-bcud}
\end{table}

\begin{table}
\centering
\caption{The predicted branching ratios for the color-suppressed processes of the type $b \rightarrow c \bar{u} d$ calculated,  in column 2, by taking $\varepsilon_1(\mu_0) = \varepsilon_8(\mu_0) = 0$ and, in column 3, by taking $\varepsilon_1(\mu_0) = -0.06 \pm 0.03, \;\;\varepsilon_8(\mu_0) = 0.12 \pm 0.02 $. The last column represents the available experimental upper bounds.}

\begin{tabular}{lccc}
& \multicolumn{3}{c}{Branching Ratio $\times 10^{-4}$} \\ \cline{2-4}
Process & Fac. & Nonfac. & Exp. \cite{ref:PDG-98} \\\hline
$\overline{B}^0 \rightarrow D^0 \pi^0$ & $0.14 \pm 0.04$ & $0.79 \pm 0.21$ & $< 1.2$ \\
$\overline{B}^0 \rightarrow D^{*0} \pi^0$ & $0.19 \pm 0.05$ & $1.11 \pm 0.30$ & $< 4.4$ \\
$\overline{B}^0 \rightarrow D^0 \eta$ & $0.08 \pm 0.02$ & $0.44 \pm 0.12$ & $< 1.3$ \\
$\overline{B}^0 \rightarrow D^{*0} \eta$ & $0.11 \pm 0.03$ & $0.61 \pm 0.16$ & $< 2.6$ \\
$\overline{B}^0 \rightarrow D^0 \eta'$ & $0.02 \pm 0.01$ & $0.13 \pm 0.04$ & $< 9.4$ \\
$\overline{B}^0 \rightarrow D^{*0} \eta'$ & $0.03 \pm 0.01$ & $0.18 \pm 0.05$ & $< 14$ \\
$\overline{B}^0 \rightarrow D^0 \rho^0$ & $0.09 \pm 0.02$ & $0.54 \pm 0.14$ & $< 3.9$ \\
$\overline{B}^0 \rightarrow D^{*0} \rho^0$ & $0.20 \pm 0.05$ & $1.16 \pm 0.31$ & $< 5.6$ \\
$\overline{B}^0 \rightarrow D^0 \omega$ & $0.09 \pm 0.02$ & $0.53 \pm 0.14$ & $< 5.1$ \\
$\overline{B}^0 \rightarrow D^{*0} \omega$ & $0.20 \pm 0.05$ & $1.15 \pm 0.31$ & $< 7.4$ \\
\end{tabular}

\label{tab:B0-CS-bcud}
\end{table}

\begin{table}
\centering
\caption{The predicted branching ratios for the color-favored processes of the type
$b \rightarrow c \bar{c} s$ calculated,  in column 2, by taking $\varepsilon_1(\mu_0) = \varepsilon_8(\mu_0) = 0$ and, in column 3, by taking $\varepsilon_1(\mu_0) = -0.06 \pm 0.03, \;\;\varepsilon_8(\mu_0) = 0.12 \pm 0.02 $. The last column represents the available experimental measurements.}

\begin{tabular}{lccc}
& \multicolumn{3}{c}{Branching Ratio $\times 10^{-3}$} \\ \cline{2-4}
Process & Fac. & Nonfac. & Exp. \cite{ref:PDG-98} \\\hline
$B \rightarrow D D_s$ & $13.9 \pm 3.0$ & $11.5 \pm 2.6$ & $9.8 \pm 2.4$ \\
$B \rightarrow D D_s^*$ & $12.2 \pm 2.5$ & $10.1 \pm 2.2$ & $9.4 \pm 3.1$\\
$B \rightarrow D^* D_s$ & $7.8 \pm 1.7$ & $6.5 \pm 1.5$ & $10.4 \pm 2.8$\\
$B \rightarrow D^* D_s^*$ & $25.9 \pm 5.2$ & $21.4 \pm 4.6$ & $22.3 \pm 5.7$\\
\end{tabular}

\label{tab:B-CF-bccs}
\end{table}

\begin{table}
\centering
\caption{The predicted branching ratios for the color-suppressed processes of the type
$b \rightarrow c \bar{c} s$ calculated,  in column 2, by taking $\varepsilon_1(\mu_0) = \varepsilon_8(\mu_0) = 0$ and, in column 3, by taking $\varepsilon_1(\mu_0) = -0.06 \pm 0.03, \;\;\varepsilon_8(\mu_0) = 0.12 \pm 0.02 $. The last column represents the available experimental measurements.}

\begin{tabular}{lccc}
& \multicolumn{3}{c}{Branching Ratio $\times 10^{-4}$} \\ \cline{2-4}
Process & Fac. & Nonfac. & Exp. \cite{ref:PDG-98,ref:CDF-98} \\\hline
$B \rightarrow K J/\psi$ & $1.7 \pm 0.3$ & $9.6 \pm 1.9$ & $9.5 \pm 0.8$ \\
$B \rightarrow K \psi(2S)$ & $0.9 \pm 0.2$ & $5.1 \pm 1.0$ & $5.8 \pm 1.2$ \\
$B \rightarrow K^* J/\psi$ & $2.7 \pm 0.5$ & $15.5 \pm 3.0$ & $14.6 \pm 1.4$ \\
$B \rightarrow K^* \psi(2S)$ & $1.6 \pm 0.3$ & $9.3 \pm 1.9$ & $9.6 \pm 2.5$ \\
\end{tabular}

\label{tab:B-CS-bccs}
\end{table}

\begin{table}
\centering
\caption{The predicted branching ratios for the decays of $B^-$ meson of the type $b \rightarrow c \bar{u} d$ calculated,  in column 2, by taking $\varepsilon_1(\mu_0) = \varepsilon_8(\mu_0) = 0$ and, in column 3, by taking $\varepsilon_1(\mu_0) = -0.06 \pm 0.03, \;\;\varepsilon_8(\mu_0) = 0.12 \pm 0.02 $. The last column represents the available experimental measurements.}

\begin{tabular}{lccc} 
& \multicolumn{3}{c}{Branching Ratio $\times 10^{-3}$} \\ \cline{2-4}
Process & Fac. & Nonfac. & Exp. \cite{ref:PDG-98} \\\hline
$B^- \rightarrow D^0 \pi^-$ & $5.1 \pm 0.2$ & $5.3 \pm 0.4$ & $5.3 \pm 0.5$ \\
$B^- \rightarrow D^0 \rho^-$ & $11.4 \pm 0.3$ & $10.7 \pm 0.8$ & $13.4 \pm 1.8$ \\
$B^- \rightarrow D^{*0} \pi^-$ & $4.3 \pm 0.1$ & $4.5 \pm 0.4$ & $4.6 \pm 0.4$ \\
$B^- \rightarrow D^{*0} \rho^-$ & $11.0 \pm 0.3$ & $11.3 \pm 0.9$ &  $15.5 \pm 3.1$\\
\end{tabular}

\label{tab:Bm-III-bcud}
\end{table}

\begin{table}
\centering
\caption{The predicted branching ratios for the color-favored $B_s$ decays of the type $b \rightarrow c \bar{u} d$ calculated,  in column 2, by taking $\varepsilon_1(\mu_0) = \varepsilon_8(\mu_0) = 0$ and,  in column 3, by taking $\varepsilon_1(\mu_0) = -0.06 \pm 0.03, \;\;\varepsilon_8(\mu_0) = 0.12 \pm 0.02 $. The last column represents the available experimental limits.}

\begin{tabular}{lccc}
& \multicolumn{3}{c}{Branching Ratio $\times 10^{-3}$} \\ \cline{2-4}
Process & Fac. & Nonfac. & Exp. \cite{ref:PDG-98} \\\hline
$\overline{B}_s \rightarrow D_s^+ \pi^-$ & $3.6 \pm 0.2$ & $3.0 \pm 0.3$ & $<130$ \\
$\overline{B}_s \rightarrow D_s^+ \rho^-$ & $8.6 \pm 0.4$ & $7.2 \pm 0.6$ & $-$ \\
$\overline{B}_s \rightarrow D_s^+ a_1^-$ & $9.8 \pm 1.0$ & $8.1 \pm 1.0$ & $-$ \\
$\overline{B}_s \rightarrow D_s^{*+} \pi^-$ & $2.6 \pm 0.1$ & $2.2 \pm 0.2$ &  $-$\\
$\overline{B}_s \rightarrow D^{*+} \rho^-$ & $7.5 \pm 0.4$ & $6.2 \pm 0.5$ & $-$\\
$\overline{B}_s \rightarrow D^{*+} a_1^-$ & $10.7 \pm 1.1$ & $8.9 \pm 1.1$  &  $-$ \\
\end{tabular}

\label{tab:Bs-CF-bcud}
\end{table}

\begin{table}
\centering
\caption{The predicted branching ratios for the color-suppressed $B_s$ decays of the type
$b \rightarrow c \bar{u} d$ calculated,  in column 2, by taking $\varepsilon_1(\mu_0) = \varepsilon_8(\mu_0) = 0$ and,  in column 3, by taking $\varepsilon_1(\mu_0) = -0.06 \pm 0.03, \;\;\varepsilon_8(\mu_0) = 0.12 \pm 0.02 $.}

\begin{tabular}{lccc}
& \multicolumn{3}{c}{Branching Ratio $\times 10^{-4}$} \\ \cline{2-4}
Process & Fac. & Nonfac. & Exp. \\\hline
$\overline{B}_s \rightarrow D^0 K^0$ & $0.18 \pm 0.05$ & $1.0 \pm 0.3$ & $-$ \\
$\overline{B}_s \rightarrow D^{*0} K^0$ & $0.25 \pm 0.07$ & $1.5 \pm 0.4$ & $-$ \\
$\overline{B}_s \rightarrow D^0 K^{*0}$ & $0.13 \pm 0.03$ & $0.7 \pm 0.2$ & $-$ \\
$\overline{B}_s \rightarrow D^{*0} K^{*0}$ & $0.27 \pm 0.07$ & $1.5 \pm 0.4$ & $-$ \\
\end{tabular}

\label{tab:Bs-CS-bcud}
\end{table}

\begin{table}
\centering
\caption{The predicted branching ratios for the color-suppressed $B_s$ decays of the type
$b \rightarrow c \bar{c} s$ calculated,  in column 2, by taking $\varepsilon_1(\mu_0) = \varepsilon_8(\mu_0) = 0$ and,  in column 3, by taking $\varepsilon_1(\mu_0) = -0.06 \pm 0.03, \;\;\varepsilon_8(\mu_0) = 0.12 \pm 0.02 $. The last column represents the available experimental measurements.}

\begin{tabular}{lccc}
& \multicolumn{3}{c}{Branching Ratio $\times 10^{-4}$} \\ \cline{2-4}
Process & Fac. & Nonfac. & Exp. \cite{ref:PDG-98} \\\hline
$\overline{B}_s \rightarrow \eta J/\psi$ & $0.44 \pm 0.08$ & $2.5 \pm 0.5$ & $< 38$ \\
$\overline{B}_s \rightarrow \eta \psi(2S)$ & $0.24 \pm 0.05$ & $1.4 \pm 0.3$ & $-$ \\
$\overline{B}_s \rightarrow \eta' J/\psi$ & $0.52 \pm 0.09$ & $2.9 \pm 0.6$ & $-$ \\
$\overline{B}_s \rightarrow \eta' \psi(2S)$ & $0.23 \pm 0.04$ & $1.3 \pm 0.3$ & $-$ \\
$\overline{B}_s \rightarrow \phi J/\psi$ & $1.83 \pm 0.32$ & $10.5 \pm 2.1$ & $9.3 \pm 3.3$ \\
$\overline{B}_s \rightarrow \phi \psi(2S)$ & $1.25 \pm 0.23$ & $7.2 \pm 1.5$ & $-$ \\
\end{tabular}

\label{tab:Bs-CS-bccs}
\end{table}

\begin{table}
\centering
\caption{Predictions of longitudinal ($P_0$) and transverse ($P_-$) polarizations for Cabibbo-favored decays of the form $B \rightarrow V V$ where $V$ is a vector meson.}

\begin{tabular}{lccccc}
& \multicolumn{2}{c}{$P_0$} && \multicolumn{2}{c}{$P_-$} \\ \cline{2-3} \cline{5-6}
Process & Fac. & Exp. &&   Fac. & Exp. \\\hline
$\overline{B}^0 \rightarrow D^{*+} \rho^-$ & 0.87 & $0.93 \pm 0.07$ \cite{ref:PDG-98}
&& 0.12 & $-$ \\
$\overline{B}^0 \rightarrow D^{*+} a_1^-$ & 0.73 & $-$ &&  0.24 & $-$ \\
$\overline{B}^0 \rightarrow D^{*0} \rho^0$ & 0.57 & $-$ && 0.43 & $-$ \\
$\overline{B}^0 \rightarrow D^{*0} \omega$ & 0.57 & $-$ && 0.43 & $-$ \\\hline
$B \rightarrow D^* D_s^*$ & 0.49 & $-$ && 0.43 & $-$ \\
$B \rightarrow K^* J/\psi$ & 0.36 & $0.52 \pm 0.08$ \cite{ref:CLEO-97} & & 0.63 & $0.47 \pm 0.08$ \cite{ref:CLEO-97,ref:Shamali-98} \\
$B \rightarrow K^* \psi(2S)$ & 0.29 & $-$ && 0.69 & $-$ \\\hline
$B^- \rightarrow D^{*0} \rho^-$ & 0.86 & $-$ &&  0.12 & $-$ \\\hline
$\overline{B}_s \rightarrow D^{*+} \rho^-$ & 0.87 &$-$ && 0.11 & $-$ \\
$\overline{B}_s \rightarrow D^{*+} a_1^-$ & 0.73 & $-$ &&  0.24 & $-$ \\
$\overline{B}_s \rightarrow D^{*0} K^{*0}$ & 0.57 & $-$ && 0.43 & $-$ \\
$\overline{B}_s \rightarrow \phi J/\psi$ & 0.35 & $0.56 \pm 0.21$ \cite{ref:CDF-95}
&& 0.64 & $-$ \\
$\overline{B}_s \rightarrow \phi \psi(2S)$ & 0.31 & $-$ && 0.66 & $-$ \\
\end{tabular}

\label{tab:B-Polarization}
\end{table}

\end{document}